\documentclass{icrc2009}

\usepackage{graphicx}   
\usepackage[caption=false]{caption}    
\usepackage[font=footnotesize]{subfig} 
\usepackage{fixltx2e}
\usepackage{url}

\newcommand{\shorttitle}[1]%
{\markboth{Proceedings of the 31\MakeLowercase{$^{st}$} ICRC, {\L}\'{o}d\'{z} 2009}{#1} }
\newcommand{\etal}{\MakeLowercase{\textit{et al. }}} 

\begin{document}
\title{Galactic diffuse gamma-ray flux at the energy about 175 TeV}

\author{\IEEEauthorblockN{Romen M. Martirosov\IEEEauthorrefmark{1},
			  Samvel V. Ter-Antonyan\IEEEauthorrefmark{2},
                       Anatoly D. Erlykin\IEEEauthorrefmark{3},
                          Alexandr P. Garyaka\IEEEauthorrefmark{1},\\
                            Natalya M. Nikolskaya\IEEEauthorrefmark{3},
                             Yves A. Gallant\IEEEauthorrefmark{4} and
                                Lawrence W. Jones\IEEEauthorrefmark{5}}
                            \\
\IEEEauthorblockA{\IEEEauthorrefmark{1}Yerevan Physics Institute, Yerevan, Armenia}
\IEEEauthorblockA{\IEEEauthorrefmark{2}Department of Physics, Southern University, Baton Rouge, LA, USA}
\IEEEauthorblockA{\IEEEauthorrefmark{3}
Russian Academy of Sciences, Lebedev Physical Institute,
Moscow, Russia}
\IEEEauthorblockA{\IEEEauthorrefmark{4}
Lab. de Physique Theorique et Astroparticules,
                         Universite Montpellier II, Montpellier, FRANCE}
\IEEEauthorblockA{\IEEEauthorrefmark{5}
Department of Physics, University of Michigan, Ann Arbor, USA}}

\shorttitle{R.M. Martirosov\etal Galactic diffuse $\gamma$-ray flux }
\maketitle

\parindent=15pt
\begin{abstract}
 We present an upper limit of galactic diffuse gamma-ray flux,
measured with the GAMMA experiment at energy about 175 TeV.
 The results were obtained using selection of muon poor extensive
air showers at mountain level (700 g/cm$^2$, Mt. Aragats, Armenia) for
5 GeV muon energy threshold. An upper limit for the differential flux
at energy $E_{\gamma}\simeq175^{+25}_{-20}$ TeV is equal to
$(5.8-7.0)\cdot10^{-12}$ $(erg\cdot m^2\cdot s\cdot sr)^{-1}$
for $95\%$ confidence level
\footnote{Corresponding author:\\
E-mail: samvel\_terantonyan@subr.edu (S.V. Ter-Antonyan)}.
  \end{abstract}


\section{Introduction}
Ultra-high energy ($E>100$ TeV) galactic gamma-radiation is an important
source of information about the origin of cosmic rays and their propagation.
 According to conventional model of cosmic rays the expected flux of
ultra-high energy $\gamma$-rays in the energy range of $0.1-1$ PeV
has presumably to have hadronic origin from sources distributed within
radii $\sim1-0.01$  Mpc \cite{Felix} respectively.\\
\indent
In the very-high (TeV) energy region $\gamma$-ray flux was measured
by ground-based systems and single Cherenkov telescopes (Whipple \cite{VER},
CANGAROO \cite{CAN}, MAGIC \cite{MAG}, H.E.S.S. \cite{HESS}, MILAGRO \cite{MIL}).
The measurements in the ultra-high  energy region ($E>100$ TeV) are still poor
and made only by extensive air showers (EAS) arrays at Chacaltaya \cite{CHAC},
MSU \cite{MSU}, Tien-Shan \cite{TS}, CASA-MIA \cite{CASA}, EAS-TOP
\cite{EASTOP}, KASCADE \cite{KAS} and Grapes-III \cite{GR}.\\
\indent
This paper is devoted to measurements of the diffuse gamma-rays with
GAMMA EAS array \cite{GAMMA,GAMMA1} at Aragats mountain observatory,
 where correlation of observable shower parameters with primary energy
is about $0.98$.  The primary nuclei (predominantly $H$)
and $\gamma$-showers were discriminated using no-muon
signal from underground muon scintillation carpet.
\\
\section{GAMMA experiment}
GAMMA experiment \cite{GAMMA,GAMMA1} is the ongoing study of primary
 energy spectra in the range of $10^{14}-10^{17}$ eV using EAS array
 with 33 concentric disposed scintillator stations (each of 3$\times$1m$^2$)
and underground scintillation carpet with 150 scintillators (each of 1m$^2$)
 to detect the shower muon component with energy $E_{\mu}>5GeV/\cos\theta$,
where $\theta$ is the shower zenith angle.
Layout of GAMMA facility is presented in Fig.~1.
\begin{figure}[!t]
  \centering
  \includegraphics[width=3.0in, height=1.5in]{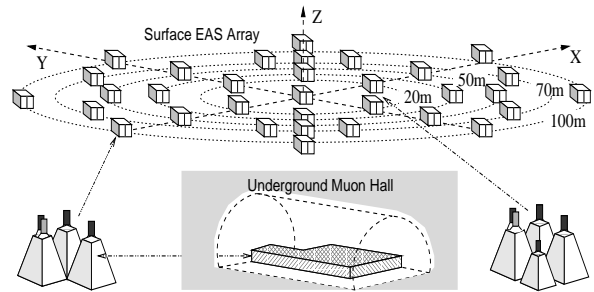}
  \caption{Layout of GAMMA experiment.}
  \label{fig1}
 \end{figure}\\
\indent
The detailed description of experiment, results of exploration of
$p, He, O$-like, and $Fe$-like primary nuclei energy spectra derived
from parametrized EAS inverse problem solution are presented in \cite{GAMMA}.
 The all-particle primary energy spectrum obtained from event-by-event
 analysis is published in \cite{GAMMA1}. \\

\indent
Here, the discrimination of $\gamma$-showers from primary nuclei induced
showers is performed on the basis of following 6 conditions:\\
1) the reconstructed shower core coordinates is distributed within
radius of $R<15$m;\\
2) shower zenith angles $\theta<30^0$;\\
3) reconstructed shower size $N_{ch}>10^5$;\\
4) reconstructed shower age parameters ($s$) is distributed within $0.4<s<1.5$;\\
5) goodness-of-fit test for reconstructed showers $\chi^2<2.5$;\\
6) no-muon signal is recorded for detected showers satisfying
the previous 5 conditions.\\
\indent
The selection criteria and $\gamma$-shower discrimination rule (6) above
 were obtained using CORSIKA shower simulation code \cite{CORSIKA}
for the NKG and EGS modes in the frameworks of the SIBYLL \cite{SIBYLL}
interaction model.
Simulations were done for 4 nuclear species: $p, He, O, Fe$ using
united energy spectral index $\gamma=-2.7$ \cite{GAMMA}.
Simulated samples were equal $7.5\times10^5$, $10^5$, $7\times10^4$
and $5\times10^4$ for $p,He,O,Fe$ nuclei and NKG mode of CORSIKA.
The samples for the EGS mode of CORSIKA were equal to $2.5\times10^4$
for primary $\gamma$-quanta, $7.5\times10^4$ for primary protons and
$3\times5000$ for $He,0$ and $Fe$ primary nuclei.
The simulation strategy and reconstruction method for shower size
($N_{ch}$), age parameter ($s$), core coordinates
($x_0,y_0$)  and shower zenith angle ($\theta$) were the same as
in \cite{GAMMA}.\\
\indent
The shower trigger efficiency and shower size reconstruction errors
($\Delta_{N_{ch}}$ and $\sigma_{N_{ch}}$) are presented in Figs.~2,3
respectively. The observed differences of reconstructed shower size
biases $\Delta_{N_{ch}}$ for different primary
particles (Fig.~3, lower panel) stems from differences of
corresponding lateral distribution functions of shower particles.
\\
\begin{figure}[!t]
  \centering
  \includegraphics[width=3.0in]{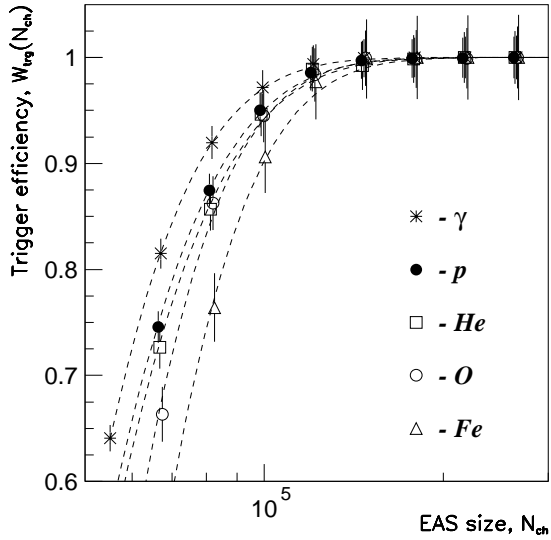}
  \caption{Trigger efficiency of GAMMA EAS array
   for different primary particles. Dashed lines
   correspond to the exponential approximations:
$W_{trg}(A,N_{ch})=1-\alpha_A\exp{(-N_{ch}/N_{0,A})}$,
where $\alpha_A$ and $N_{0,A}$ parameters depend on primary 
particle ($A$).}
\label{fig2}
 \end{figure}
\begin{figure}[!t]
  \centering
  \includegraphics[width=3.0in]{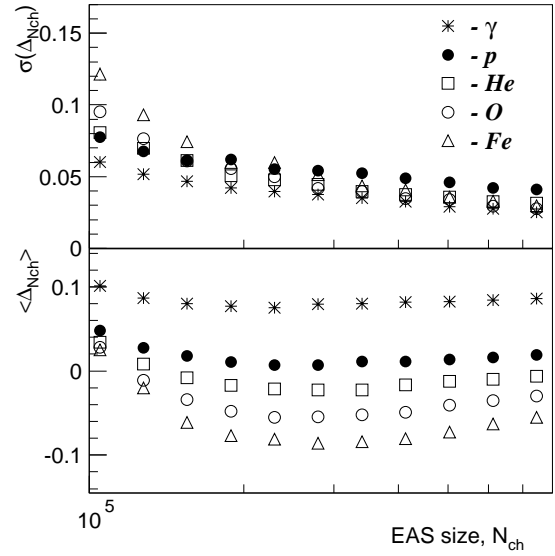}
  \caption{Expected reconstruction error (upper panel) and average bias
(lower panel) of shower size ($N_{ch}$) for different primary particles
(symbols). $\Delta_{ch}=\ln (N_{ch}^*/N_{ch})$ and $N_{ch}^*$ is an
estimation of shower size $N_{ch}$.}
  \label{fig3}
 \end{figure}
\indent
The distribution of detected shower age parameters (GAMMA data)
in comparison with expected distributions for primary $p,He,O,Fe$ nuclei
are presented in Fig.~4 (left panel). The primary elemental
\begin{figure}[!t]
  \centering
  \includegraphics[width=3.0in]{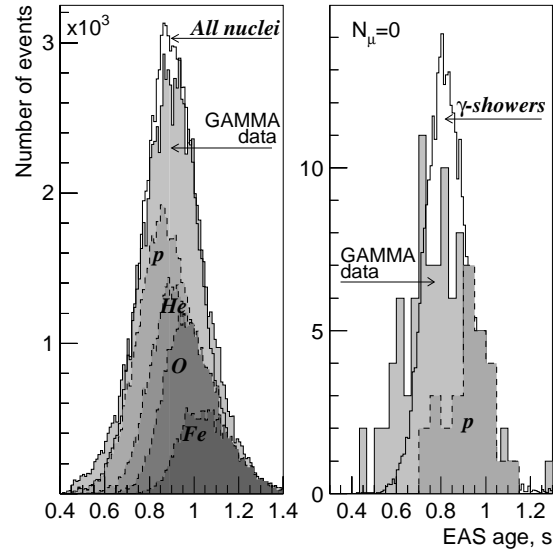}
\caption{Shower age parameter ($s$) distribution for all showers
(left panel) and no-muon detected showers ($N_{\mu}=0$, right
panel). Simulated data for the primary mixed composition
($All$ $nuclei$, left panel) and primary gamma ray (right panel)
are normalized to the corresponding GAMMA experimental data.}
  \label{fig4}
 \end{figure}
composition and energy spectra were taken
from solution of parametrized EAS inverse problem \cite{GAMMA}
in the frameworks of the SIBYLL interaction model (Fig.~5,
 shaded area \cite{GAMMA}) and were extrapolated up to the
100 TeV energy region. The reliability  of this extrapolations
stems from data  \cite{Sineg}.\\
\begin{figure}[!t]
  \centering
  \includegraphics[width=3.0in]{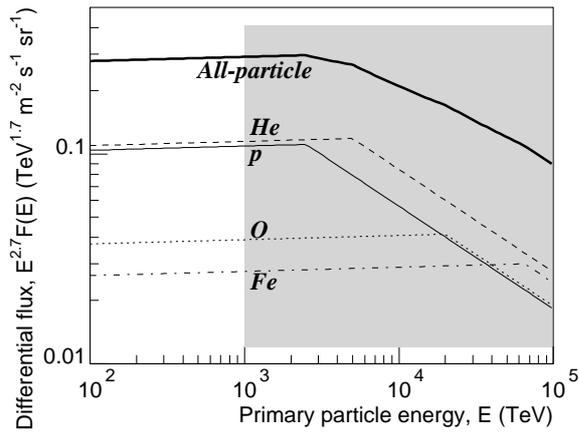}
\caption{Primary energy spectra and all-particle energy spectrum
taken from \cite{GAMMA} (shaded area) and corresponding
extrapolations to the 100 TeV energy region.}
  \label{fig5}
 \end{figure}
\indent
The right panel of  Fig.~4 shows distribution of shower age parameters
for selected no-muon signal events (shades GAMMA data area)
in comparison with corresponding expected distributions from
the simulated
$\gamma$-showers and background proton showers.
It is well seen, that the EAS age parameter is also a data carrier
about primary particle ($\gamma$-showers are younger). However,
 we did not include yet the age parameter in the $\gamma$-showers
 selection criteria. Results in Fig.~4 we use only as indication
 of an agreement between simulated and corresponding
detected distributions.\\
\indent
The detected muon number spectra in the normalization of probability
density function for different shower size thresholds
($N_{ch}>10^5, 2\times10^5$ and $4\times10^5$) are presented in Fig.~6
\begin{figure}[!t]
  \centering
  \includegraphics[width=3.0in]{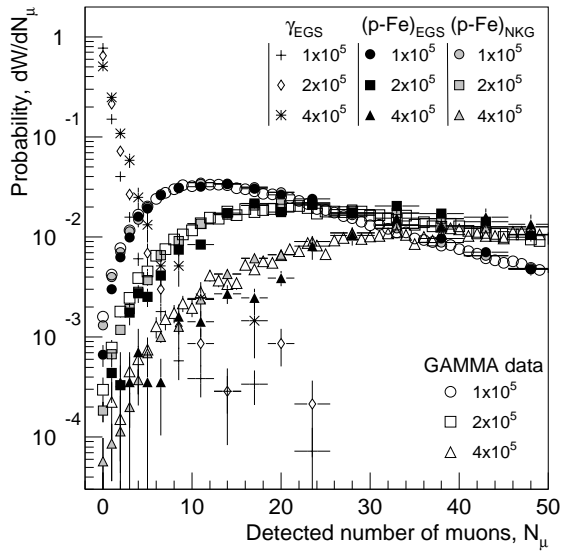}
\caption{Normalized detected muon number ($N_\mu$) spectra for
 different shower size
thresholds ($10^5,2\times10^5,4\times10^5$). Hollow symbols
(circle, square and triangle) are GAMMA experimental data. The symbols in
$\gamma_{EGS}$, $(p-Fe)_{EGS}$ and $(p-Fe)_{NKG}$ columns
correspond to simulated
data for the primary $\gamma$ and mixed composition
($p,He,O,Fe$ \cite{GAMMA}) computed using the EGS and NKG modes of CORSIKA.}
  \label{fig6}
 \end{figure}
 (hollow symbols) in comparison with the corresponding expected spectra
from different primary particles ($\gamma,p,He,O,Fe$) and different
simulation modes (NKG, EGS) of CORSIKA. Energy spectra and elemental
composition of primary nuclei (Fig.~5) used in the Figs.~4,6 were taken
 from \cite{GAMMA} and applied for the energy region $E>100$ TeV.
\\
\subsection{Energy estimation}
The energy of primary particle is estimated using event-by-event method
\cite{GAMMA1} according to the empirical expression:
$\ln{E_A}=A_1\ln{N_{ch}}+A_2/\cos{\theta}+A_3$,
where $E$ is in GeV, parameters $A_1,A_2$ and $A_3$ are determined
using goodness-of-fit test for simulated database
and depend on primary particle $A$.\\
\indent
The corresponding accuracies providing for $\chi^2\simeq1$ is described
by the log-linear functions\\
$\sigma(E_A)=\varepsilon_A-\delta_A\ln{(E_A/10^5)}$,
where $\varepsilon\equiv0.22$, $0.30$, $0.33$ and
$\delta\equiv0.01$, $0.02$, $0.05$
for primary $\gamma$, $p$ and proton induced no-muon detected
shower ($p_0$) respectively.\\
\indent
The primary energy reconstruction
efficiencies are presented in Fig.~7 for different
primary particles.
The lines represent the log-linear empiric expression above for primary
 proton showers (dashed line) and $\gamma$-showers (solid line)
respectively. It is seen, that the proton produced no-muon signal
 events (hollow circles) practically similar to $\gamma$-showers.
 The inset histogram shows the distribution of $E_{p,\gamma}/N_{ch}$ ratio
for $p$ and $\gamma$ primary particles respectively.\\
\begin{figure}[!t]
  \centering
  \includegraphics[width=3.0in]{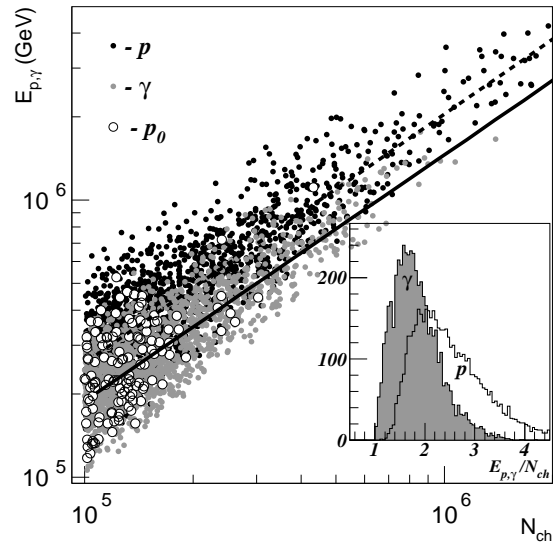}
\caption{Primary energy ($E$) and corresponding shower size ($N_{ch}$)
distributions at observation level for 5000 primary protons (bold dot
symbols) and 5000 $\gamma$ (gray dot symbols). Hollow circles correspond to
the proton showers with no-muon signal from underground muon carpet
($N_{\mu}=0$).
Solid and dashed lines are the log-linear approximations (see text) for primary
$\gamma$ and $p$ correspondingly. Inset histograms are
$E_{p,\gamma}/N_{ch}$ distributions for primary proton and $\gamma$
(shaded area).}
  \label{fig7}
 \end{figure}

\section{Gamma ray flux}
$98000$ shower events were selected for operation time
$T=3970$ h of GAMMA experiment. Number of detected
shower events versus number of detected muons ($N_{\mu}$) for different
shower size thresholds $N_{ch}>10^5,2\times10^5$ and $4\times10^5$ are
presented in Fig.~8 (histograms with shaded area).
The symbols in Fig.~8 are the corresponding
expected number of events
simulated using the CORSIKA code for primary energy spectra
\cite{GAMMA} presented in Fig.~5. The simulations were carried out for
two modes of CORSIKA to get high accuracy of simulation (EGS mode) and
large simulated sample (NKG mode).\\
\indent
The agreement of simulated and
detected muon spectra in all measurement region and lack of
statistically significant excess of no-muon signal events (Fig.~8)
allowed us to estimate only an upper limit of $\gamma$-ray flux
according to the expression
\begin{figure}[!t]
  \centering
  \includegraphics[width=3.0in]{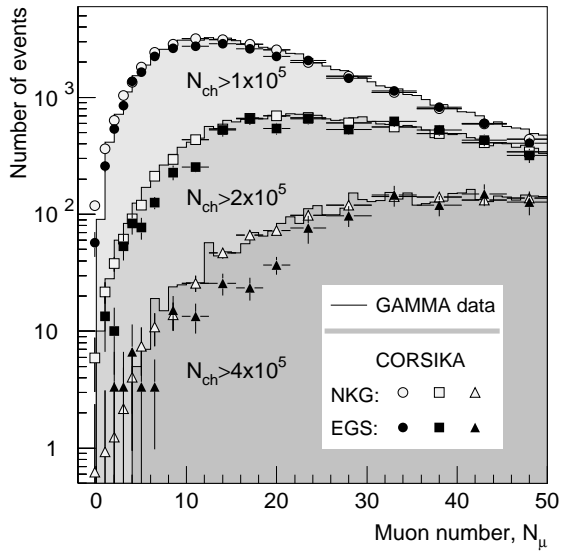}
\caption{Detected (histogram lines) and expected (symbols)
muon number ($N_{\mu}$) spectra for different shower size thresholds
($10^5,2\times10^5,4\times10^5$) and different mode (NKG, EGS) of CORSIKA.}
  \label{fig8}
 \end{figure}
\begin{equation}
J=\frac{2\sqrt{M_0/(W_{\gamma,N_{\mu}=0}W_{trg,\gamma})}}
{S\Omega T\overline{\cos{\theta}}}\cdot\frac{1}{\Delta E}
\end{equation}
where $M_0$ is the number of no-muon detected showers,
 $W_{\gamma,N_{\mu}=0}$ is the probability
to detect no-muon signal for $\gamma$-showers (see Fig.~6),
$W_{trg}(E_{\gamma})$ is the trigger efficiency (see Fig.~2),
$\overline{\cos{\theta}}=0.94$ is the average
shower zenith angle, $S$ and $\Omega$ are the EAS core
detection area and corresponding solid angle.\\
\indent
The obtained upper limit of differential $\gamma$-ray flux in the
energy range 100-300 PeV
are presented in Fig.~9 (black downward triangle symbol) in comparison
with CASA-MIA \cite{CASA}, KASCADE \cite{KAS} and EAS-TOP
\cite{EASTOP} data.
\begin{figure}[!t]
  \centering
  \includegraphics[width=3.0in]{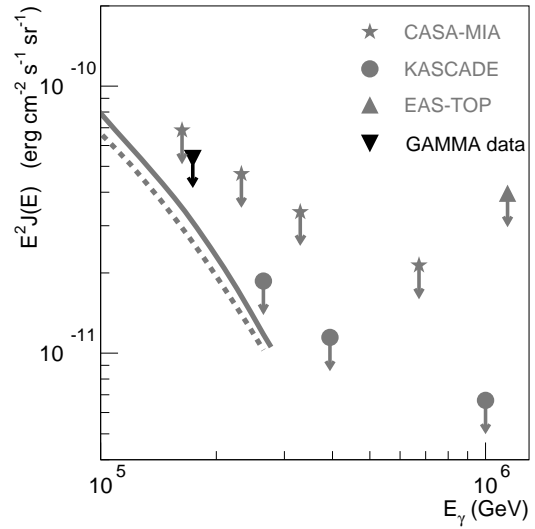}
\caption{Upper limit of gamma ray flux derived from detected no-muon
showers (black downward triangle symbol).
The gray symbols are the CASA-MIA \cite{CASA}, KASCADE
\cite{KAS} and EAS-TOP \cite{EASTOP} data taken from
\cite{KAS}. The lines are expected Galactic diffuse background
flux from \cite{AA}.}
  \label{fig9}
 \end{figure}
\\

\section{Conclusion}
An upper limit of $\gamma$-ray differential flux at energy
$E_\gamma\simeq175^{+25}_{-20}$ TeV obtained with GAMMA experiment
is equal to
$(5.8-7.0)\cdot10^{-12}$ $(erg\cdot m^2\cdot s\cdot sr)^{-1}$
for $95\%$ confidence level
and it is in close agreement with the CASA-MIA data \cite{CASA}.\\
\indent
The lower limit for the primary energy spectra and
elemental composition obtained with the GAMMA experiment
\cite{GAMMA} can be
extended to the lower energy region up to about 100 TeV energies.\\
\indent
We are going to increase the underground muon carpet area up to
250 m$^2$ to improve $\gamma$/proton showers discrimination efficiency.
\\
\section{Acknowledgment}
This work has been partly supported by the research grant no. 090 from
the Armenian government, the RFBR grant 07-02-00491 in Russia, by the
Moscow P.N.~Lebedev Physical Institute and the Hayastan All-Armenian
Fund.
\\

\end{document}